\documentclass[aps,prd,preprint,onecolumn,showpacs,
superscriptaddress,nofootinbib,amsmath,amssymb]{revtex4-1}

\usepackage{graphicx,color}
\ifdefined\XeTeXversion
  \DeclareGraphicsExtensions{.pdf,.png,.jpg}
\else
  \ifdefined\pdfoutput
    \ifnum\pdfoutput>0
      \DeclareGraphicsExtensions{.pdf,.png,.jpg}
    \else
      \DeclareGraphicsExtensions{.eps}
    \fi
  \else
    \DeclareGraphicsExtensions{.eps}
  \fi
\fi
\newcommand{\rev}[1]{{\color{black}#1}}
\usepackage[utf8]{inputenc}
\usepackage[T1]{fontenc}

\def\re#1{Re(#1)}
\def\im#1{Im(#1)}

\begin{document}
\title{Regular black hole sourced by the Dehnen-type distribution of matter: The sound of the event horizon}
\author{Erdin\c{c} Ula\c{s} Saka}
\email{ulassaka@istanbul.edu.tr}
\affiliation{Department of Physics, Faculty of Science, Istanbul University, Vezneciler, 34134 Istanbul, T\"urkiye}

\begin{abstract}
We compute the fundamental and overtone quasinormal modes of a regular, asymptotically flat black hole supported by a Dehnen-type matter halo. Gravitational perturbations in this background split into two distinct axial sectors, and \rev{our analysis confirms that the presence of the halo parameter $a$ separates the spectra of the two axial matter-perturbation prescriptions, which coincide in vacuum.} The dependence of the quasinormal spectrum on $a$ is moderate for the fundamental modes and even weaker for the overtones, which approach one another in the complex-frequency plane as the halo parameter increases. \rev{We have found that  no pronounced overtone outburst is observed at least for the first several overtones,} indicating that the halo does not induce the type of strong near-horizon effects characteristic of quantum-corrected or exotic compact objects. Overall, our results show that the dark-matter halo introduces controlled and comparatively mild modifications to the ringdown spectrum while preserving its qualitative structure.
\end{abstract}

\maketitle

\section{Introduction}

Quasinormal modes (QNMs) form the backbone of black-hole spectroscopy, governing the characteristic ringdown phase following a dynamical perturbation \cite{Kokkotas:1999bd,  Berti:2009kk, Konoplya:2011qq, Bolokhov:2025uxz}. In recent years, the detection of gravitational-wave signals from compact-object mergers has elevated QNMs from a theoretical construct to a precision observational tool, capable of probing the strong-gravity regime of general relativity and constraining possible deviations from vacuum black-hole spacetimes \cite{LIGOScientific:2016aoc, LIGOScientific:2017vwq, LIGOScientific:2020zkf}. While much of the existing literature has historically focused on the fundamental QNMs, it has become increasingly clear that the higher overtones play an important  role in the early-time ringdown signal \cite{Giesler:2019uxc,Mitman:2025hgy,Konoplya:2022pbc,Konoplya:2023hqb,Giesler:2024hcr,Konoplya:2022iyn,Konoplya:2022hll}.

This realization was emphasized in seminal studies \cite{Giesler:2019uxc} demonstrating that overtones can substantially improve the modeling of gravitational-wave data and extend the regime in which ringdown templates accurately describe numerical and observational waveforms \cite{Giesler:2019uxc,Mitman:2025hgy}. In particular, it has been shown that the inclusion of several overtones allows one to reconstruct the post-merger signal at times much closer to the peak of radiation, thereby enhancing the potential for precision tests of black-hole geometry and dynamics. At the same time, these works have highlighted that overtones possess qualitatively different properties compared to the fundamental mode, including stronger sensitivity to the details of the effective potential and, in particular, to near-horizon modifications of the spacetime \cite{Konoplya:2022pbc}.

Motivated by these developments, a growing number of studies have explored how overtones behave in non-vacuum and non-Schwarzschild backgrounds, including black holes surrounded by matter/fields distributions, modified-gravity corrections, or regular cores replacing the classical singularity \cite{Konoplya:2025hgp,Zhu:2024wic,Stuchlik:2025ezz,Konoplya:2024lch,Bolokhov:2023bwm,Konoplya:2023aph,Konoplya:2026cus,Bolokhov:2023ruj,Lutfuoglu:2025ljm,Tang:2024txx, Konoplya:2024kih,Zhang:2024nny,Konoplya:2024hfg,Gong:2023ghh,Siqueira:2025lww,Konoplya:2023kem,Konoplya:2023ppx}. Regular black holes embedded in astrophysical environments could be of particular interest in this context, although the literature in this setting is usually limited by the least damped modes \cite{Lutfuoglu:2025kqp,Pathrikar:2025sin,Chakraborty:2024gcr,Zhang:2021bdr,Rincon:2025buq,Dubinsky:2025fwv,Pezzella:2024tkf, Malik:2025czt,Zhao:2023tyo,Feng:2025iao,Hamil:2025pte,Tovar:2025apz,Mollicone:2024lxy,Konoplya:2021ube,Daghigh:2022pcr,Liu:2024bfj,Liu:2024xcd}.

A notable recent advance in modelling black holes in realistic environments is the derivation of a family of regular, asymptotically flat solutions sourced by an anisotropic fluid that effectively represents a galactic dark--matter halo~\cite{Konoplya:2025ect}. In contrast to earlier constructions---often based on numerical solutions or matter sectors incompatible with standard energy conditions---this approach provides simple analytic metrics for a broad class of density profiles, including the commonly used Dehnen distributions (see \cite{Bolokhov:2026opp,Bolokhov:2025zva} for a review). These geometries possess no curvature singularities and interpolate smoothly between a Schwarzschild-like exterior and an inner de~Sitter core, all while satisfying the weak energy condition. They therefore offer a physically well-motivated framework for modelling supermassive black holes embedded in galactic halos and a convenient testbed for studying how environmental matter affects spacetime structure, quasinormal spectra, and related observational signatures.

In two recent works, the fundamental QNMs of the above mentioned regular black holes \cite{Konoplya:2025ect} supported by matter halos have been calculated and shown to exhibit moderate but systematic shifts relative to their vacuum counterparts \cite{Lutfuoglu:2025mqa,Bolokhov:2025fto}. These studies established the stability of the configurations and clarified the dependence of the lowest-lying frequencies on the halo parameters. However, despite the growing importance of higher overtones for gravitational-wave phenomenology, a detailed analysis of the overtone spectrum in such matter-supported regular spacetimes has remained unexplored.

The purpose of the present work is to fill this gap by systematically studying the first several quasinormal overtones of a regular black hole immersed in a galactic halo. Focusing on gravitational perturbations, we compute the complex frequencies of higher-order modes using the Leaver numerical techniques applicable beyond the fundamental ringdown. By extending the analysis to overtones, we aim to assess their sensitivity to the halo scale, explore departures from the vacuum Schwarzschild behavior, and clarify the extent to which environmental effects become more pronounced at higher damping rates. The key question is whether the astrophysical environment could produce an outburst of overtones similar to that produced by the near-horizon physics \cite{Konoplya:2022pbc}.

Our results show that, while the fundamental modes provide a robust description of the late-time ringdown, the overtones encode additional information about the surrounding matter distribution and the internal structure of the regular geometry. However, no pronounced outburst of overtones is observed even for sufficiently dense models of the halo. \rev{This statement concerns the first five overtones. Notice however, that asymptotically high overtones are not expected to be observed and usually represent purely theoretical interest, so that only the first several overtones influence the early ringdown phase.}

The paper is organized as follows. Section~\ref{sec:wavelike} introduces the background metric, reviews the construction of the axial gravitational perturbation equations, and presents the corresponding effective potentials for the ``up'' and ``down'' sectors. The numerical techniques used to compute the quasinormal spectrum are described in Sec.~\ref{sec:WKB}, where we discuss the WKB method, and Sec.~\ref{sec:Leaver}, where the Leaver (Frobenius) continued-fraction approach is adapted to the asymptotically flat geometry. Section~\ref{sec:QNMs}  contains the core results of the work: the quasinormal frequencies of the first several overtones, their dependence on the halo scale parameter, the splitting between the two axial sectors, and the interpretation of these trends in the context of regular black-hole geometries. Finally, Sec.~\ref{sec:cocl} summarizes our findings.

\section{Black-hole geometry and axial gravitational perturbations}
\label{sec:wavelike}

The spacetime considered here belongs to a recently constructed class of regular, asymptotically flat black holes supported by a Dehnen-type dark-matter halo \cite{Konoplya:2025ect}. In this model, Einstein gravity is sourced by an anisotropic fluid whose density profile reproduces the well-known Dehnen family of galactic distributions, yet the resulting geometry is obtained in closed analytic form.  The static, spherically symmetric line element reads
\begin{equation}
ds^{2}
 = -f(r)\,dt^{2}  + \frac{dr^{2}}{f(r)} + r^{2}\left(d\theta^{2}+\sin^{2}\!\theta\,d\phi^{2}\right),
\end{equation}
with metric function
\begin{equation}
f(r)=1-\frac{2M r^{2}}{(r+a)^{3}} ,
\end{equation}
where $M$ is the ADM mass and $a>0$ sets the halo scale.   The corresponding density profile may be written in the standard Dehnen form \cite{Dehnen:1993uh,Taylor:2002zd},
\begin{equation}
\rho(r)=\rho_{0} \left(\frac{r}{a}\right)^{-\alpha} \!\left(1+\frac{r^{k}}{a^{k}}\right)^{-(\gamma-\alpha)/k},
\end{equation}
with parameters $\gamma=4$, $\alpha=0$ and $k=1$ adopted in what follows. At large distances, the spacetime reduces to the Schwarzschild form,
\(
f(r)=1-2M/r + \mathcal{O}(r^{-2}),
\)
while near the center, one finds
\(
f(r)=1 - 2M r^{2}/a^{3} + \mathcal{O}(r^{3}),
\)
revealing a regular de Sitter core. Thus, the metric supplies a physically motivated model of a black hole embedded in a realistic galactic environment, free of curvature singularities and satisfying the weak energy condition.

\smallskip

Linear perturbations of such matter-supported backgrounds require additional care, since the anisotropic fluid contributes nontrivially to the perturbed Einstein equations. A recent analysis \cite{Chakraborty:2024gcr} showed that the axial sector can be consistently formulated in two gauge-invariant ways, depending on whether the \emph{contravariant} or \emph{covariant} components of the fluid variables are kept unperturbed.  These two frameworks lead to distinct master equations, which we refer to as the ``up'' and ``down'' formulations.
Both reduce to the standard Regge--Wheeler equation in the vacuum limit, but they probe different combinations of the background stress--energy components.

\rev{The two spectra correspond to different assumptions about perturbations of the anisotropic matter: holding contravariant and covariant fluid variables unperturbed imposes inequivalent restrictions when the metric is perturbed \cite{Chakraborty:2024gcr,Konoplya:2025ect}. Both equations have axial (odd) parity. Their separation therefore compares two axial matter-response prescriptions, rather than axial and polar spectra; no polar modes are calculated here.}

\smallskip

After the usual tensor-harmonic decomposition and the Regge--Wheeler redefinition of the axial amplitude, the perturbations satisfy a Schr\"odinger-type master equation,
\begin{equation}\label{masterwave}
\frac{d^{2}\Psi}{dr_{*}^{2}} + \bigl[\omega^{2} - V(r)\bigr]\Psi = 0, \qquad \frac{dr_{*}}{dr}=\frac{1}{f(r)} ,
\end{equation}
where $r_{*}$ is the tortoise coordinate mapping the exterior region to $(-\infty,+\infty)$.

\smallskip

The effective potentials obtained in \cite{Chakraborty:2024gcr} take the form
\begin{eqnarray}\nonumber
V^{\mathrm{up}}(r) &=& f(r)\!\left[ \frac{\ell(\ell+1)}{r^{2}} -\frac{6m(r)}{r^{3}} + 4\pi\bigl(\rho -5P_{r}+4P\bigr)\right], \\
V^{\mathrm{down}}(r) &=& f(r)\!\left[ \frac{\ell(\ell+1)}{r^{2}}  -\frac{6m(r)}{r^{3}} + 4\pi\bigl(\rho - P_{r}\bigr) \right],
\end{eqnarray}
where $m(r)$ is the Misner--Sharp mass and $\rho$, $P_{r}$ and $P$ denote the energy density, radial pressure, and tangential pressure of the halo. For the present background, the matter variables satisfy
\begin{equation}
P_{r} = -\rho,
\qquad
P = -\rho - \frac{r}{2}\rho'(r),
\end{equation}
which simplifies the above expressions to
\begin{eqnarray}\nonumber
V^{\mathrm{up}}(r)
 &=& f(r)\!\left[ \frac{\ell(\ell+1)}{r^{2}} -\frac{6m(r)}{r^{3}} + 8\pi\rho - 8\pi r\rho'(r)\right], \\[1mm]
V^{\mathrm{down}}(r)  &=& f(r)\!\left[ \frac{\ell(\ell+1)}{r^{2}} -\frac{6m(r)}{r^{3}} + 8\pi\rho \right].
\end{eqnarray}

Both potentials are positive-definite outside the horizon and display a single barrier shape, ensuring mode stability. Because the matter sector enters differently in the ``up'' and ``down'' constructions, the two potentials generally produce distinct quasinormal spectra---a feature that does not occur in vacuum and will play a central role in our analysis of gravitational ringdown and wave propagation.

\section{WKB method}\label{sec:WKB}

For asymptotically flat black holes whose effective potential forms a single, smooth barrier, the quasinormal spectrum can be approximated semi-analytically by the WKB expansion~\cite{Iyer:1986np,Konoplya:2003ii,Matyjasek:2017psv}. Applied to the master equation \ref{masterwave}, the method expresses the complex frequencies in terms of derivatives of the potential at its maximum $r_{0}$.

Quasinormal modes are defined as solutions of the perturbation equation that satisfy purely ingoing behaviour at the event horizon and purely outgoing radiation at spatial infinity. In terms of the tortoise coordinate $r_{*}$, which maps the exterior region to $(-\infty,+\infty)$, these conditions read
\begin{equation}\label{qnmbc}
\begin{array}{rcll}
\Psi(r_{*}) &\propto& e^{-i\omega r_{*}}, &\qquad r_{*}\to -\infty \quad \text{(ingoing at the horizon)},\\
\Psi(r_{*}) &\propto& e^{+i\omega r_{*}}, &\qquad r_{*}\to +\infty \quad \text{(outgoing at infinity)}.
\end{array}
\end{equation}
No incoming radiation from spatial infinity is allowed, and no signal emerges from the horizon. These two boundary conditions uniquely define the discrete quasinormal spectrum for massless perturbations of asymptotically flat black holes.

In the standard notation,
\begin{equation}\label{eq:WKBbasic-short}
\omega^{2}=V_{0}+\Lambda_{2}+\Lambda_{4}+\cdots - i\Bigl(n+\tfrac12\Bigr)\sqrt{-2V_{2}}\,   \left(1+\Lambda_{3}+\Lambda_{5}+\cdots\right),
\end{equation}
where $V_{i}\equiv d^{i}V/dr_{*}^{i}|_{r_{0}}$ and the $\Lambda_{k}$ denote the WKB correction terms. This expansion is asymptotic, and its convergence strongly depends on the potential; therefore Pad\'e resummation of the truncated series is commonly employed to improve stability and accuracy~\cite{Bolokhov:2022rqv,Skvortsova:2024atk,Lutfuoglu:2025blw,del-Corral:2022kbk,Skvortsova:2023zmj,Zhao:2022gxl,Kodama:2009bf,Bolokhov:2025egl,Skvortsova:2024wly,Lutfuoglu:2025pzi,Skvortsova:2025cah,Bolokhov:2025lnt,Skvortsova:2024msa,Dubinsky:2025wns,Zinhailo:2018ska}.
A rational Pad\'e approximant,
\[
P_{k/m}\{\omega^{2}\}
   =\frac{\sum_{i=0}^{k}a_{i}(\omega^{2})^{i}}
          {1+\sum_{j=1}^{m}b_{j}(\omega^{2})^{j}},
\]
provides a significantly more accurate estimate than the raw WKB series.

In our calculations, we use the sixth- and ninth-order expansions (WKB$_6$ and WKB$_9$) together with Pad\'e approximants (typically $P_{3/3}$ or $P_{4/5}$). \rev{The spread between these two orders is an internal order-sensitivity diagnostic and is quoted separately from the actual Frobenius discrepancy in the QNM tables of Sec.~\ref{sec:QNMs}. It is not a reliable bound on the true error, particularly outside the low-overtone WKB regime.} This strategy has proved accurate for a broad class of black-hole perturbation problems, including scalar, electromagnetic, Dirac, and gravitational fields~\cite{Bolokhov:2023bwm,Konoplya:2023ahd,Bolokhov:2024ixe,Albuquerque:2023lhm,Lutfuoglu:2025eik,Ishihara:2008re,Bonanno:2025dry,Guo:2022hjp,Zhidenko:2003wq,Skvortsova:2024eqi,Konoplya:2022hbl,Bolokhov:2023ruj,Abdalla:2005hu,Paul:2023eep,Arbelaez:2025gwj,Lutfuoglu:2025qkt,Konoplya:2001ji,Kokkotas:2010zd,Lutfuoglu:2025bsf,Zinhailo:2019rwd}.

Although highly effective for the fundamental mode and low overtones ($n<\ell$), the WKB+Pad\'e method becomes less reliable for very high $n$, for potentials with multiple peaks, or when the eikonal structure is strongly modified.
Within its expected regime of validity, however, it offers an efficient and accurate tool for determining quasinormal frequencies of the present regular black-hole geometry. Here, we will use the WKB method as a complementary approach, while the precise Leaver method will be used as the main method for finding overtones.

\rev{In order to estimate the error of the WKB method, we define the relative difference between the k-th order WKB value and precise value obtained by the Frobenius method, 
\begin{equation}\label{eq:benchmarkerror}
\delta_k=100\frac{|\omega_{{\rm WKB},k}-\omega_{\rm F}|}{|\omega_{\rm F}|},\qquad k=6,9.
\end{equation}
The retained inter-order difference is $\Delta_{69}=100|\omega_6-\omega_9|/|\omega_9|$. Agreement between approximants may reflect a common bias, so $\Delta_{69}$ and the measured discrepancies $\delta_k$ have different meanings.}

\section{Leaver (Frobenius) method}
\label{sec:Leaver}

For asymptotically flat black holes, QNMs can be computed with very high precision using the Frobenius-based method introduced by Leaver \cite{Leaver:1985ax,Leaver:1990zz}. In contrast to semi-analytic techniques such as the WKB approximation, Leaver's method yields essentially exact results (up to controllable numerical accuracy). It is therefore particularly well-suited for benchmarking approximate methods and for calculating higher overtones.

We begin with the master perturbation equation, which, in terms of the radial coordinate $r$ is a second-order ordinary differential equation with regular singular points at the event horizon $r=r_{+}$ and an irregular singular point at spatial infinity $r=\infty$. To impose the quasinormal boundary conditions \eqref{qnmbc}, we define the new function, $y(r)$, such that
\begin{equation}\label{reg}
\Psi(r) = e^{i\omega r} r^{\sigma} \left(1 - \frac{r_{+}}{r}\right)^{-i\omega / f'(r_{+})} y(r),
\end{equation}
where the exponent $\sigma$ is determined by substituting Eq.~\eqref{reg} into the wave equation \eqref{masterwave} and expanding in powers of $1/r$. With this choice, the quasinormal boundary conditions are satisfied provided that $y(r)$ is regular at both $r=r_{+}$ and $r=\infty$.

We expand $y(r)$ in a Frobenius series around the horizon,
\begin{equation}\label{Frobenius}
y(r) = \sum_{i=0}^{\infty} a_i \left(1 - \frac{r_{+}}{r}\right)^i.
\end{equation}
Substituting this expansion into Eq.~\eqref{masterwave} yields \rev{a finite-term linear recurrence relation for the coefficients}
\begin{equation}\label{eq:n-recurrence}
c_{0,i}a_i+c_{1,i}a_{i-1}+c_{2,i}a_{i-2}+c_{3,i}a_{i-3}+\ldots=0,
\end{equation}
which can be reduced to a three-term recurrence relation by Gaussian elimination (see \cite{Konoplya:2011qq} for details),
\begin{equation}
\alpha_{i} a_{i+1}+\beta_{i} a_{i}+\gamma_{i} a_{i-1}=0, \qquad i\ge 1,
\label{eq:recurrence}
\end{equation}

The quasinormal frequencies follow from the requirement that the Frobenius series \eqref{Frobenius} converges at spatial infinity. This condition leads to an infinite continued-fraction equation for $\omega$,
\begin{equation}
\beta_{0} - \frac{\alpha_{0}\gamma_{1}}{\beta_{1} - \displaystyle\frac{\alpha_{1}\gamma_{2}}{\beta_{2} - \displaystyle\frac{\alpha_{2}\gamma_{3}}{\beta_{3}-\cdots}}}=0. \label{eq:continued_fraction}
\end{equation}
We solve Eq.~\eqref{eq:continued_fraction} numerically in the complex $\omega$-plane. To accelerate convergence of the continued fraction, we employ the Nollert improvement \cite{Nollert:1993zz}, extended in \cite{Zhidenko:2006rs} to handle recurrence relations with an arbitrary number of terms.

The Leaver method typically converges very rapidly and yields highly accurate QNM spectra for asymptotically flat black holes, including high overtones. For this reason, it is widely used as a reference standard against which approximate methods---such as WKB expansions---are tested \cite{Benda:2025tni,Konoplya:2004uk,Dias:2022oqm,Bolokhov:2024bke,Stuchlik:2025mjj,Xiong:2023usm,Kanti:2006ua,Zinhailo:2024kbq,Konoplya:2007zx,Konoplya:2017tvu,Onozawa:1996ux}.  When additional singular points lie between the horizon and spatial infinity, we employ continuation of the Frobenius expansion through a sequence of positive real midpoints, following \cite{Rostworowski:2006bp}.

In the present work, Leaver's method is employed to obtain precise quasinormal frequencies for massless fields in asymptotically flat geometries and to assess the reliability of semi-analytic approaches, especially for higher overtones where WKB-based techniques are known to lose accuracy.

\rev{Tables~\ref{tableI} and~\ref{tableII} compare the Frobenius values directly with WKB for the fundamental mode and fifth overtone, respectively. Tables~\ref{tab:frobenius_down} and~\ref{tab:frobenius_up} tabulate both axial spectra for all studied $n=0,\ldots,5$ at representative parameters. The full frequency lists are supplied with the numerical data.}

\begin{table*}
\centering\color{black}
\renewcommand{\baselinestretch}{1}\footnotesize
\renewcommand{\arraystretch}{1.05}
\setlength{\tabcolsep}{4pt}
\begin{tabular*}{\linewidth}{@{\extracolsep{\fill}}rccc}
\hline\hline
$a/r_+$ & Frobenius & WKB$_6$ ($P_{3/3}$) & WKB$_9$ ($P_{4/5}$) \\
\hline
$0.00$ & $0.7473434-0.1779246i$ & $0.7472397-0.1778656i$ & $0.7473176-0.1779467i$\\
$0.04$ & $0.7362133-0.1626553i$ & $0.7362073-0.1626551i$ & $0.7362134-0.1626558i$\\
$0.08$ & $0.7224870-0.1485987i$ & $0.7224933-0.1485920i$ & $0.7224871-0.1485989i$\\
$0.12$ & $0.7061355-0.1353912i$ & $0.7061312-0.1353729i$ & $0.7061356-0.1353912i$\\
$0.16$ & $0.6871575-0.1228322i$ & $0.6871510-0.1228280i$ & $0.6871576-0.1228320i$\\
$0.20$ & $0.6655643-0.1108389i$ & $0.6655580-0.1108352i$ & $0.6655651-0.1108394i$\\
$0.24$ & $0.6413804-0.0994179i$ & $0.6413778-0.0994190i$ & $0.6413805-0.0994177i$\\
$0.28$ & $0.6146555-0.0886480i$ & $0.6146628-0.0886524i$ & $0.6146556-0.0886476i$\\
$0.32$ & $0.5854906-0.0786718i$ & $0.5855024-0.0786540i$ & $0.5854907-0.0786712i$\\
$0.36$ & $0.5540852-0.0696845i$ & $0.5540802-0.0696750i$ & $0.5540852-0.0696842i$\\
$0.40$ & $0.5208049-0.0619018i$ & $0.5208009-0.0618975i$ & $0.5208048-0.0619016i$\\
$0.44$ & $0.4862392-0.0554827i$ & $0.4862363-0.0554798i$ & $0.4862391-0.0554827i$\\
$0.48$ & $0.4511843-0.0504290i$ & $0.4511818-0.0504266i$ & $0.4511841-0.0504291i$\\
\hline\hline
\multicolumn{4}{c}{Relative differences (\%)}\\
$a/r_+$ & Inter-order $\Delta_{69}$ & Actual $\delta_6$ & Actual $\delta_9$ \\
\hline
$0.00$ & $0.014633$ & $0.015524$ & $0.004421$\\
$0.04$ & $0.000815$ & $0.000803$ & $0.00006$\\
$0.08$ & $0.001261$ & $0.00125$ & $0.000024$\\
$0.12$ & $0.00262$ & $0.002621$ & $0.000007$\\
$0.16$ & $0.001104$ & $0.001102$ & $0.000026$\\
$0.20$ & $0.001224$ & $0.001087$ & $0.000137$\\
$0.24$ & $0.000452$ & $0.00043$ & $0.000033$\\
$0.28$ & $0.00139$ & $0.00137$ & $0.000076$\\
$0.32$ & $0.00352$ & $0.003619$ & $0.000106$\\
$0.36$ & $0.00187$ & $0.001931$ & $0.000067$\\
$0.40$ & $0.001083$ & $0.001121$ & $0.000038$\\
$0.44$ & $0.000823$ & $0.000842$ & $0.000031$\\
$0.48$ & $0.000756$ & $0.000766$ & $0.000036$\\
\hline\hline\end{tabular*}
\caption{Quasinormal modes of the $\ell=2$ up potential for the Konoplya--Zhidenko black hole at $r_+=1$, $n=0$. Frequencies are $r_+\omega$. The original inter-order difference is retained as the diagnostic $\Delta_{69}$; the actual errors $\delta_6$ and $\delta_9$ are measured against Frobenius frequencies as defined in Eq.~\eqref{eq:benchmarkerror}. All up frequencies use the stated density-gradient contribution $-8\pi r\rho\prime(r)$.}\label{tableI}
\end{table*}

\begin{table*}
\centering\color{black}
\renewcommand{\baselinestretch}{1}\footnotesize
\renewcommand{\arraystretch}{1.05}
\setlength{\tabcolsep}{4pt}
\begin{tabular*}{\linewidth}{@{\extracolsep{\fill}}rccc}
\hline\hline
$a/r_+$ & Frobenius & WKB$_6$ ($P_{3/3}$) & WKB$_9$ ($P_{4/5}$) \\
\hline
$0.00$ & $0.3385988-2.3912161i$ & $0.3615749-2.5322845i$ & $0.1800336-2.3752078i$\\
$0.04$ & $0.4403997-2.1330297i$ & $0.4726296-2.1963064i$ & $0.4185726-2.1290623i$\\
$0.08$ & $0.4905172-1.9038280i$ & $0.4891491-1.9375512i$ & $0.5385278-1.8690581i$\\
$0.12$ & $0.5174209-1.6957854i$ & $0.5332427-1.7158790i$ & $0.4952328-1.6330430i$\\
$0.16$ & $0.5290047-1.5061370i$ & $0.5641826-1.5201894i$ & $0.5253879-1.4663810i$\\
$0.20$ & $0.5285373-1.3330761i$ & $0.5443014-1.3340462i$ & $0.5439702-1.3354013i$\\
$0.24$ & $0.5173287-1.1753734i$ & $0.5112789-1.1690308i$ & $0.5261270-1.1749617i$\\
$0.28$ & $0.4953709-1.0327409i$ & $0.4808257-1.0267689i$ & $0.5129742-1.0379282i$\\
$0.32$ & $0.4621345-0.9072377i$ & $0.4484786-0.9021960i$ & $0.4760434-0.8934791i$\\
$0.36$ & $0.4208392-0.8039922i$ & $0.4111203-0.7953354i$ & $0.4265967-0.7938359i$\\
$0.40$ & $0.3806142-0.7229624i$ & $0.3724138-0.7121504i$ & $0.3842050-0.7219546i$\\
$0.44$ & $0.3460453-0.6577263i$ & $0.3371597-0.6489376i$ & $0.3543408-0.6604053i$\\
$0.48$ & $0.3170821-0.6033170i$ & $0.3086392-0.5956980i$ & $0.3243251-0.6025424i$\\
\hline\hline
\multicolumn{4}{c}{Relative differences (\%)}\\
$a/r_+$ & Inter-order $\Delta_{69}$ & Actual $\delta_6$ & Actual $\delta_9$ \\
\hline
$0.00$ & $10.07813$ & $5.918141$ & $6.599031$\\
$0.04$ & $3.976288$ & $3.260396$ & $1.018573$\\
$0.08$ & $4.341009$ & $1.716726$ & $3.015188$\\
$0.12$ & $5.340827$ & $1.442499$ & $3.753603$\\
$0.16$ & $4.258646$ & $2.372978$ & $2.500734$\\
$0.20$ & $0.096741$ & $1.101367$ & $1.088334$\\
$0.24$ & $1.241966$ & $0.682551$ & $0.685876$\\
$0.28$ & $2.939283$ & $1.372749$ & $1.602202$\\
$0.32$ & $2.855666$ & $1.429723$ & $1.921525$\\
$0.36$ & $1.725355$ & $1.434236$ & $1.286511$\\
$0.40$ & $1.875077$ & $1.660899$ & $0.456481$\\
$0.44$ & $2.756205$ & $1.681608$ & $1.172947$\\
$0.48$ & $2.501029$ & $1.66857$ & $1.068758$\\
\hline\hline\end{tabular*}
\caption{Quasinormal modes of the $\ell=2$ up potential for the Konoplya--Zhidenko black hole at $r_+=1$, $n=5$. Frequencies are $r_+\omega$. The original inter-order difference is retained as the diagnostic $\Delta_{69}$; the actual errors $\delta_6$ and $\delta_9$ are measured against Frobenius frequencies as defined in Eq.~\eqref{eq:benchmarkerror}. All up frequencies use the stated density-gradient contribution $-8\pi r\rho\prime(r)$.}\label{tableII}
\end{table*}

\begin{figure*}
\resizebox{\linewidth}{!}{\includegraphics{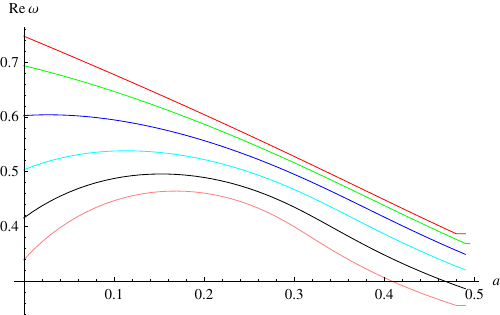}\includegraphics{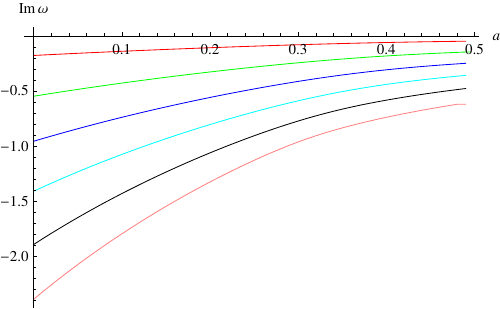}}
\caption{\rev{Quasinormal modes found by the Leaver method for the down channel: the fundamental mode and first five overtones as a function of the regularization parameter $a$, at $\ell=2$, $r_+=1$, and $n=0,\ldots,5$. The left and right panels show $\operatorname{Re}\omega$ and $\operatorname{Im}\omega$, respectively.}}\label{fig:L2Down}
\end{figure*}

\begin{figure*}
\resizebox{\linewidth}{!}{\includegraphics{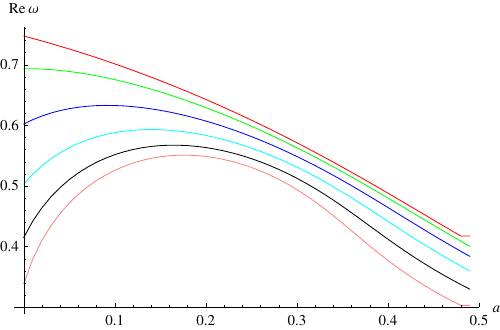}\includegraphics{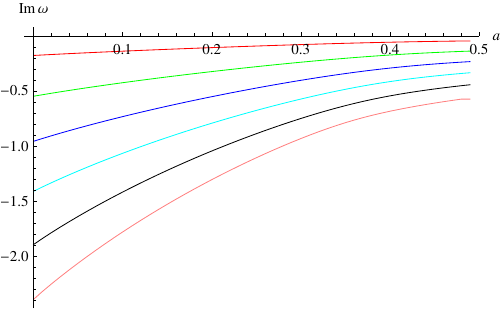}}
\caption{\rev{Leaver frequency plots for the up channel: the fundamental mode and first five overtones as a function of $a$, at $\ell=2$, $r_+=1$, and $n=0,\ldots,5$. The left and right panels show $\operatorname{Re}\omega$ and $\operatorname{Im}\omega$. The curves retain the originally supplied up-sector data before the density-gradient correction; the corrected frequencies are tabulated in Table~\ref{tab:frobenius_up}.}}\label{fig:L2Up}
\end{figure*}

\begin{figure*}
\resizebox{\linewidth}{!}{\includegraphics{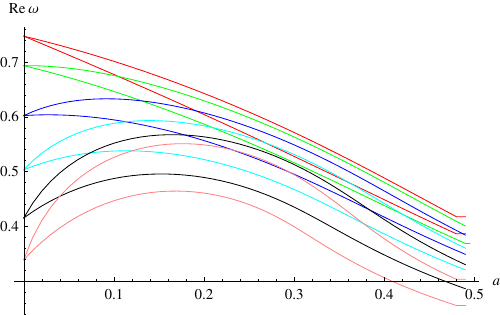}\includegraphics{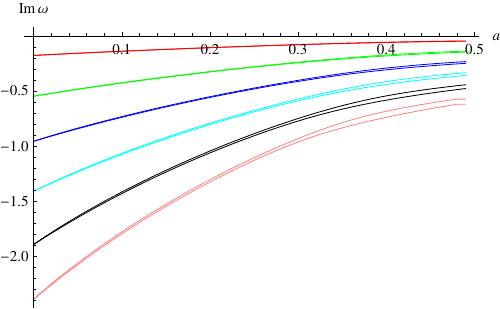}}
\caption{\rev{Leaver frequency plots for both up and down axial matter-response prescriptions, at $\ell=2$, $r_+=1$, and $n=0,\ldots,5$, as functions of $a$. The left and right panels show $\operatorname{Re}\omega$ and $\operatorname{Im}\omega$. The spectral separation compares two axial prescriptions. }}\label{fig:L2}
\end{figure*}

\rev{Figures~\ref{fig:L2Down}, \ref{fig:L2Up}, and~\ref{fig:L2} show the frequency trajectories for the down potential, the up potential, and their comparison, respectively. The effective potentials are positive definite, implying the stability of  perturbations, that is, absence of modes growing in time.}

\section{Quasinormal modes}\label{sec:QNMs}

\begin{table*}
\centering
\renewcommand{\arraystretch}{1.18}
\setlength{\arrayrulewidth}{0.45pt}
\color{black}\small
\setlength{\tabcolsep}{4pt}
\begin{tabular*}{\linewidth}{@{\extracolsep{\fill}}rccc}
\hline\hline
$A$ & $n=0$ & $n=1$ & $n=2$\\\hline
$0.00$ & $0.7473434-0.1779246i$ & $0.6934220-0.5478298i$ & $0.6021069-0.9565540i$\\
$0.10$ & $0.6772545-0.1401440i$ & $0.6463307-0.4287486i$ & $0.5943272-0.7394975i$\\
$0.20$ & $0.6040993-0.1077288i$ & $0.5865369-0.3276322i$ & $0.5564769-0.5592234i$\\
$0.30$ & $0.5275571-0.0807475i$ & $0.5160378-0.2442932i$ & $0.4949135-0.4133280i$\\
$0.40$ & $0.4484222-0.0603657i$ & $0.4377456-0.1823294i$ & $0.4166452-0.3081058i$\\
$0.44$ & $0.4168739-0.0543765i$ & $0.4062285-0.1643966i$ & $0.3851850-0.2784256i$\\
\hline\hline
$A$ & $n=3$ & $n=4$ & $n=5$\\\hline
$0.00$ & $0.5030099-1.4102964i$ & $0.4150292-1.8936898i$ & $0.3385988-2.3912161i$\\
$0.10$ & $0.5373743-1.0761717i$ & $0.4871517-1.4316425i$ & $0.4465018-1.7970128i$\\
$0.20$ & $0.5218723-0.8049194i$ & $0.4890815-1.0620088i$ & $0.4605430-1.3261217i$\\
$0.30$ & $0.4670989-0.5895475i$ & $0.4349362-0.7728625i$ & $0.3991287-0.9627352i$\\
$0.40$ & $0.3858643-0.4406798i$ & $0.3474870-0.5834724i$ & $0.3058905-0.7389527i$\\
$0.44$ & $0.3547292-0.3996934i$ & $0.3175136-0.5315244i$ & $0.2782107-0.6756762i$\\
\hline\hline\end{tabular*}
\caption{Frobenius frequencies $r_+\omega$ of the \textbf{down} axial prescription for $\ell=2$ and all computed overtone numbers $n=0,\ldots,5$ at representative $A=a/r_+$. The fundamental mode is $n=0$. The event-horizon radius is fixed to $r_+=1$, with $M/r_+=(1+A)^3/2$; these are not fixed-mass frequencies. Each entry is a calculated sample, without interpolation. Seven decimal places are displayed; the numerical checks are described in the text.}\label{tab:frobenius_down}
\end{table*}

\begin{table*}
\centering
\renewcommand{\arraystretch}{1.18}
\setlength{\arrayrulewidth}{0.45pt}
\color{black}\small
\setlength{\tabcolsep}{4pt}
\begin{tabular*}{\linewidth}{@{\extracolsep{\fill}}rccc}
\hline\hline
$A$ & $n=0$ & $n=1$ & $n=2$\\\hline
$0.00$ & $0.7473434-0.1779246i$ & $0.6934220-0.5478298i$ & $0.6021069-0.9565540i$\\
$0.10$ & $0.7146399-0.1419046i$ & $0.6863577-0.4334357i$ & $0.6389757-0.7453603i$\\
$0.20$ & $0.6655643-0.1108389i$ & $0.6494986-0.3361787i$ & $0.6216988-0.5710520i$\\
$0.30$ & $0.6003686-0.0835495i$ & $0.5898727-0.2521204i$ & $0.5703952-0.4245633i$\\
$0.40$ & $0.5208049-0.0619018i$ & $0.5109952-0.1864362i$ & $0.4914980-0.3131725i$\\
$0.44$ & $0.4862392-0.0554827i$ & $0.4762895-0.1672285i$ & $0.4564643-0.2814108i$\\
\hline\hline
$A$ & $n=3$ & $n=4$ & $n=5$\\\hline
$0.00$ & $0.5030099-1.4102964i$ & $0.4150292-1.8936898i$ & $0.3385988-2.3912161i$\\
$0.10$ & $0.5872298-1.0811590i$ & $0.5419815-1.4345482i$ & $0.5062082-1.7973684i$\\
$0.20$ & $0.5889305-0.8173226i$ & $0.5569299-1.0726392i$ & $0.5285373-1.3330761i$\\
$0.30$ & $0.5442972-0.6018599i$ & $0.5137539-0.7833922i$ & $0.4801414-0.9675663i$\\
$0.40$ & $0.4625848-0.4436661i$ & $0.4249628-0.5796830i$ & $0.3806142-0.7229624i$\\
$0.44$ & $0.4271270-0.3999310i$ & $0.3894373-0.5248554i$ & $0.3460453-0.6577263i$\\
\hline\hline\end{tabular*}
\caption{Frobenius frequencies $r_+\omega$ of the \textbf{up} axial prescription for $\ell=2$ and all computed overtone numbers $n=0,\ldots,5$ at representative $A=a/r_+$. The fundamental mode is $n=0$. The event-horizon radius is fixed to $r_+=1$, with $M/r_+=(1+A)^3/2$; these are not fixed-mass frequencies. Each entry is a calculated sample, without interpolation. Seven decimal places are displayed; the numerical checks are described in the text.}\label{tab:frobenius_up}
\end{table*}

\begin{table*}
\centering
\renewcommand{\arraystretch}{1.18}
\setlength{\arrayrulewidth}{0.45pt}
\color{black}\small\setlength{\tabcolsep}{3pt}
\begin{tabular*}{\linewidth}{@{\extracolsep{\fill}}lrccr}
\hline\hline
Prescription & $a/M$ & Frobenius $M\omega$ & Published WKB$_9$ $M\omega$ & Difference (\%) \\
\hline
down & $0.00$ & $0.37367168-0.08896232i$ & $0.37365880-0.08897340i$ & $0.004425$ \\
down & $0.10$ & $0.41979716-0.09203626i$ & $0.41979690-0.09203660i$ & $0.0001008$ \\
down & $0.20$ & $0.48998054-0.09374041i$ & $0.48998060-0.09374060i$ & $0.0000401$ \\
\hline
up & $0.00$ & $0.37367168-0.08896232i$ & $0.37365880-0.08897340i$ & $0.004425$ \\
up & $0.10$ & $0.43397486-0.09257436i$ & $0.42522380-0.09160900i$ & $1.984$ \\
up & $0.20$ & $0.52979682-0.09579707i$ & $0.50925370-0.09293880i$ & $3.852$ \\
\hline\hline\end{tabular*}
\caption{Fundamental-mode comparison at $\ell=2$, $n=0$, with Tables~I and~III of Ref.~\cite{Lutfuoglu:2025mqa}. Both sets are expressed in mass units and evaluated at the same $a/M$. The present Frobenius frequencies use the explicit potentials~of Sec.~\ref{sec:wavelike}; the published column is the ninth-order Pad\'e result, retained at its stated precision. The final column is $100|\omega_{\rm published}-\omega_{\rm F}|/|\omega_{\rm F}|$, calculated before rounding. The nonzero-parameter Frobenius points are calculated directly, without interpolation. The down-sector agreement and the up-sector discrepancy are discussed separately.}\label{tab:prior_comparison}
\end{table*}

\rev{For the comparison with Ref.~\cite{Lutfuoglu:2025mqa}, its $M=1$ normalization and our $r_+=1$ normalization are related by
\begin{equation}\label{eq:comparison_units}
 A=\frac{a}{r_+},\qquad \frac{M}{r_+}=\frac{(1+A)^3}{2},\qquad
 \frac{a}{M}=\frac{2A}{(1+A)^3},\qquad
 M\omega=\frac{(1+A)^3}{2}\,(r_+\omega).
\end{equation}
Table~\ref{tab:prior_comparison} compares the same $\ell=2$, $n=0$ geometry and prescription at $a/M=0,0.10,0.20$, without interpolation. The nonzero down-sector differences from the published ninth-order values are $0.000101\%$ and $0.0000401\%$. The corresponding up-sector differences are $1.98\%$ and $3.85\%$, which we report explicitly. 
The present up frequencies are checked by continued-fraction refinement and independent fundamental-mode calculations.
}

\rev{For the $n=5$ overtone in Table~\ref{tableII}, the actual complex-frequency discrepancies from Frobenius range from $0.683\%$ to $5.92\%$ at sixth order and from $0.456\%$ to $6.60\%$ at ninth order. These direct benchmarks assess accuracy; the separate spread between the two WKB truncations is only an order-sensitivity diagnostic, not a reliable estimate of the true error.}

This behavior is illustrated explicitly by the $n=5$ overtone of the \rev{up sector.}
The WKB method for $ a=0.48$ (see Table \ref{tableII}) gives
\[
\omega_{n=5}^{\mathrm{WKB}} = \rev{0.30863920 - 0.59569800}\, i ,
\]
whereas the Frobenius method yields a considerably different value
\[
\omega_{n=5}^{\mathrm{Frob}} = \rev{0.31708211 - 0.60331699}\, i .
\]
\rev{For this example the actual discrepancy is $1.67\%$ at sixth order and $1.07\%$ at ninth order. These differences are much larger than the fundamental-mode discrepancies, and should not be identified with an error bound inferred from the order spread.}

In contrast, for the fundamental mode, the situation is entirely different. For $n=0$,  $ a=0.48$,  we find excellent agreement between the two approaches (see Table \ref{tableI}):
\[
\omega_{n=0}^{\mathrm{WKB}} = \rev{0.45118179 - 0.05042659}\, i ,
\]
\[
\omega_{n=0}^{\mathrm{Frob}} = \rev{0.45118429 - 0.05042900}\, i ,
\]
with differences well below the percent level.
This confirms that the WKB method is highly reliable for the fundamental mode and, as also demonstrated in Ref.~\cite{Lutfuoglu:2025mqa}, for the first overtone.

We therefore conclude that, although the WKB approximation provides accurate results for the lowest-lying QNMs, it cannot be used to reliably estimate higher overtones in the present system. Since the WKB series is asymptotic rather than convergent at each fixed order, \rev{its accuracy for the sampled higher overtones must be assessed from the Frobenius data rather than inferred from order spread.} \rev{For a fixed effective potential, the excellent fundamental-mode agreement in Table~\ref{tableI} supports the use of WKB in the earlier fundamental-mode studies~\cite{Lutfuoglu:2025mqa,Bolokhov:2025fto}. This accuracy does not extend automatically to all higher overtones: percent-level discrepancies are already present at $n=5$ in Table~\ref{tableII}. Accurate determination of these overtone frequencies therefore requires a convergent numerical method with controlled truncation, such as the Frobenius (Leaver) approach used in the present work.}

In addition to the behaviour already discussed, several further features of the quasinormal spectrum deserve emphasis. First, the dependence of the frequencies on the halo parameter $a$ is not uniform across all modes: while the imaginary parts $\im{\omega_{n}}$ show an almost monotonic shift as $a$ increases, the real parts $\re{\omega_{n}}$ exhibit a clearly non-monotonic behaviour, in agreement with Figs.~\ref{fig:L2Down}--\ref{fig:L2}. Second, \rev{the ``up'' and ``down'' axial matter-response prescriptions have distinct spectra when the halo is present.}  This splitting is already visible for the fundamental mode, but it persists and remains of comparable magnitude for the higher overtones, demonstrating that the halo modifies the effective potentials in genuinely different ways.

A further noteworthy feature is that, for $a>0$, the spacing between successive overtones decreases: the modes lie closer to each other in the complex frequency plane than in the Schwarzschild case ($a=0$).  In other words, the introduction of the halo compresses the overtone spectrum.  This behaviour is mild and systematic, with no indication of the dramatic ``overtone bursts'' that can occur in scenarios involving strong near-horizon modifications or exotic effective geometries.  Thus, although the halo affects both the fundamental and excited modes, it does so smoothly, without producing large enhancements in the higher-$n$ sector.  This reinforces the interpretation that the dark-matter profile provides a gentle deformation of the Schwarzschild potential rather than introducing qualitatively new short-distance physics.

\rev{The absence of a pronounced outburst stated here is restricted to the computed $\ell=2$, $n=0,\ldots,5$ cases and the sampled nonextremal range.  However, higher multipoles are usually even less sensitive to overtones behavior, because the centrifugal barrier has dominant contribution. Thus even smaller relative  deviations of overtones is expected for higher multipoles.}

\begin{figure*}
\centering\includegraphics[width=\linewidth]{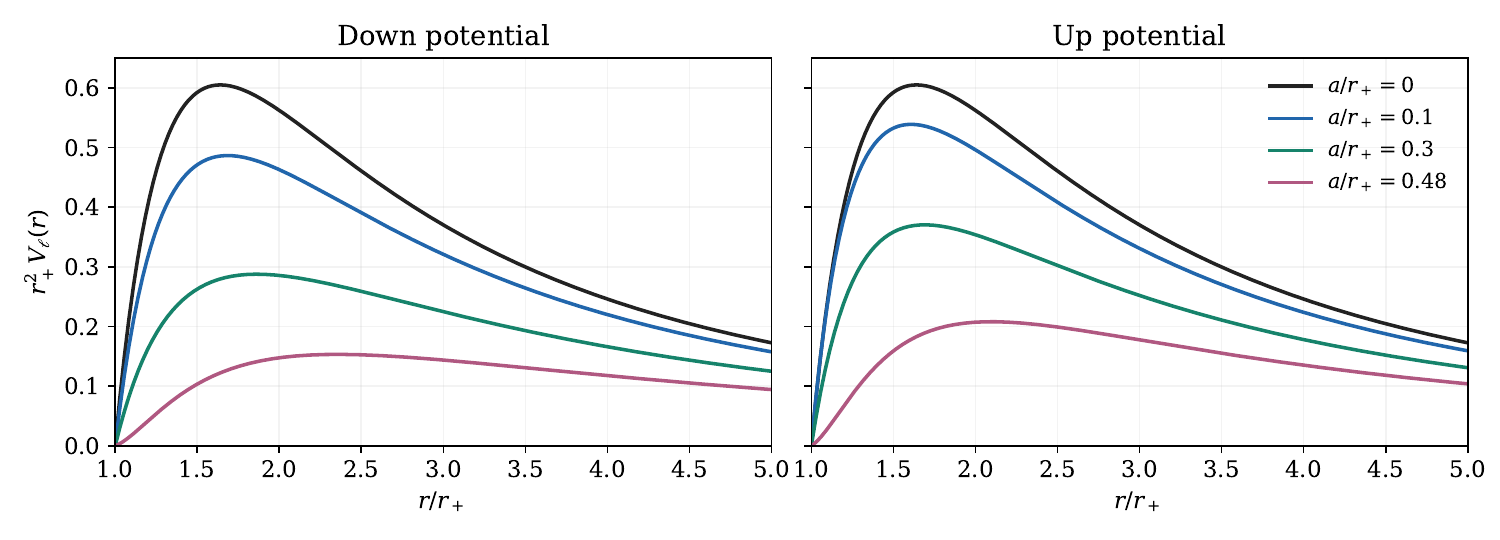}
\caption{\rev{Exterior axial potentials at $\ell=2$ for $A=a/r_+=0,0.1,0.3,0.48$, with $r_+=1$ and $M/r_+=(1+A)^3/2$. The left and right panels show the down and up prescriptions, respectively, evaluated from the potentials given in Sec.~\ref{sec:wavelike}. The vertical coordinate is $r_+^2V$. The curves retain a smooth single exterior barrier in these examples. The difference $V^{\rm up}-V^{\rm down}=24Marf/(r+a)^5$ vanishes at the horizon and infinity. }}\label{fig:potentials}
\end{figure*}

\rev{The qualitative potential comparison in Fig.~\ref{fig:potentials} illustrates the smooth exterior deformation. In the present model,
\begin{equation}\label{eq:potential_difference}
 V^{\rm up}-V^{\rm down}=\frac{24Mar f(r)}{(r+a)^5}.
\end{equation}
This matter contribution vanishes at the horizon and at infinity and changes the single exterior barrier smoothly, without introducing a secondary narrow barrier in the plotted examples. This gives a qualitative explanation for the smooth frequency trajectories observed for various overtones.

Notice that the quasinormal modes, namely, the fundamental mode and the first overtone, found here could be used for finding the other characteristic of radiation - grey-body factors \cite{Konoplya:2026qql} of gravitational field propagating in the vicinity of the black hole.
One can use the correspondence suggested in Eq.~(3.5) of Ref.~\cite{Konoplya:2024lir}:
\begin{equation}\label{eq:qnm_greybody_application}
\Gamma_\ell(\Omega)\equiv |T|^2=
\left[1+\exp\!\left(
2\pi\frac{\Omega^2-[\operatorname{Re}(\omega_0)]^2}
{4\operatorname{Re}(\omega_0)\operatorname{Im}(\omega_0)}
\right)\right]^{-1}+\mathcal{O}(\ell^{-1}).
\end{equation}
Here $\Omega$ is the real scattering frequency, $T$ is the transmission amplitude, and $\operatorname{Im}(\omega_0)<0$. This leading correspondence requires only the fundamental frequency, while the subdominant terms include also the first overtone. For the present $\ell=2$ perturbations, the up- and down-sector values in Tables~\ref{tab:frobenius_down} and~\ref{tab:frobenius_up} can be inserted separately to estimate $\Gamma_2^{\rm up}(\Omega)$ and $\Gamma_2^{\rm down}(\Omega)$.

The formula thus provides a direct application of the computed spectrum to the halo dependence of partial-wave transmission and absorption. The first-overtone frequencies tabulated here supply the additional input for higher-order refinements of the correspondence. This correspondence has been tested and used in a great number of recent publications \cite{Malik:2025dxn,Bolokhov:2026kqu,Bolokhov:2026eqf,Lutfuoglu:2025hjy,Shi:2025gst, Han:2025cal,Dubinsky:2024vbn,Lutfuoglu:2025ldc,Malik:2024cgb,Konoplya:2024vuj,Lutfuoglu:2025ohb,Bolokhov:2026uol}. Notice that the above correspondence does work when the quasinormal modes cannot be well reproduced via the WKB method, so that a number of exceptions exist, including various higher curvature corrected theories and asymptotically non-flat spacetimes \cite{Konoplya:2022gjp,Konoplya:2017wot,Bolokhov:2023dxq,Konoplya:2025afm}. Our case, however, does not belong to any of the above exceptions.}

\section{Conclusions}\label{sec:cocl}

In this work. we have computed the gravitational quasinormal spectrum of the regular Konoplya--Zhidenko black hole supported by a Dehnen-type halo, with particular emphasis on the first several overtones.  While the fundamental modes of this geometry were analysed previously in \cite{Lutfuoglu:2025mqa,Bolokhov:2025fto}, the behaviour of the higher overtones and their sensitivity to the halo parameter had not been systematically explored.

Our analysis shows that the presence of the halo parametrised by $a$ modifies all quasinormal frequencies, but the magnitude of this effect remains modest: both the real and imaginary parts of the frequencies change smoothly across the parameter range considered. The real parts of the frequencies exhibit a non-monotonic dependence on $a$, whereas the imaginary parts typically vary monotonically. An important qualitative observation is that the spacing between successive overtones becomes smaller once $a\neq 0$, so that the overtones move closer to each other in the complex-frequency plane.  This behaviour contrasts with scenarios in which near-horizon modifications or new-physics corrections produce rapidly growing deviations for higher overtones. In the present case the halo deforms the spectrum only mildly, \rev{without a pronounced outburst along the computed $n\leq5$ branches.}

Because the background contains anisotropic matter, axial perturbations split into two inequivalent sectors (``up'' and ``down''), and \rev{we have verified that these two axial matter-response prescriptions have distinct spectra.}  The splitting is visible for all computed modes, including the overtones, but remains small compared with the absolute values of the frequencies.  The deviation increases with~$a$ yet never becomes dominant, indicating that the associated matter terms modify the axial potentials in a controlled rather than drastic way.

The quasinormal frequencies were computed using the WKB method up to ninth order with Pad\'e improvement and compared with the Frobenius (Leaver) method whenever the latter converged. For the fundamental mode, the agreement between WKB and Frobenius is excellent. For higher overtones, however, the WKB method ceases to be reliable: \rev{the actual discrepancies in Tables~\ref{tableI} and~\ref{tableII} are measured directly against Frobenius results; the inter-order difference is not a reliable error bound.} \rev{For the sampled overtones, the Frobenius frequencies provide the quantitative reference; the measured WKB accuracy is that given in the benchmark tables, rather than a universal threshold in $n$.} \rev{Thus, while WKB calculations can adequately describe the fundamental mode for a given effective potential, the discrepancies already seen at $n=5$ establish the need for a convergent numerical approach, such as Frobenius, for precise overtone spectroscopy in the present model.}

Overall, our findings demonstrate that the Dehnen-supported regular black hole exhibits a stable and smoothly varying quasinormal spectrum whose dependence on the halo parameter is moderate and does not produce any dramatic overtone effects. \rev{For the computed $n\leq5$ branches, the absence of a pronounced outburst and the separation of the two axial prescriptions indicate} that environmental matter modifies the ringdown signal in a gentle rather than disruptive manner.

\begin{acknowledgments}
The author thanks Bekir Can L\"utf\"uo\u{g}lu for valuable assistance with the numerical calculations and for insightful scientific discussions during the preparation of this manuscript.
\end{acknowledgments}

\bibliography{bibliographyNEW}
\end{document}